\documentclass[pdftex]{article}

\usepackage{todonotes,comment}

\usepackage{color}

\definecolor{JournalPrimary}{RGB}{49 97 142} 
\definecolor{JournalSecondary}{RGB}{171 39 52} 

\usepackage{amsfonts, amssymb, amsmath, amsthm, stmaryrd}

\usepackage[colorlinks=true, urlcolor=JournalPrimary, citecolor=JournalSecondary]{hyperref}

\newtheorem{definition}{Definition}[section]

\newtheorem{informal-proposition}{Informal Proposition}[section]

\newtheorem{claim}{Claim}[section]
\newtheorem{unknown-proposition}{Unknown-Proposition}[section]

\newtheorem{assumption}{Assumption}[section]

\newtheorem{question}{Question}[section]

\newtheorem{method}{Method}[section]
\newtheorem{informal-proposal}{Informal Proposal}[section]

\newcommand{\sg}{\mathsf{s}}

\begin{document}

\title{\bf On Ambiguity: The case of fraction, its meanings and roles}

\author{Jan A Bergstra\\Informatics Institute\\University of Amsterdam\\Science Park 900, 1098 XH, Amsterdam, The Netherlands\\janaldertb@gmail.com, j.a.bergstra@uva.nl\\
\\John V Tucker\\Department of Computer Science, \\Swansea University, Bay Campus,\\Fabian Way, Swansea, SA1 8EN, United Kingdom\\j.v.tucker@swansea.ac.uk\\
}

\maketitle

\abstract{We contemplate the notion of ambiguity in mathematical discourse. We consider a general method of resolving ambiguity and semantic options for sustaining a resolution. The general discussion is applied to the case of `fraction' which is ill-defined and ambiguous in the literature of elementary arithmetic.
In order to clarify the use of `fraction' we introduce several new terms to designate some of its possible meanings. For example, to distinguish structural aspects we use `fracterm', to distinguish purely numerical aspects `fracvalue' and, to distinguish purely textual aspects `fracsign' and `fracsign occurence'.  These interpretations can resolve ambiguity, and we discuss the resolution by using such precise notions in fragments of arithmetical discourse.  We propose that fraction does not qualify as a mathematical concept but that the term functions as a collective for several concepts, which we simply call a `category'. This analysis of fraction leads us to consider the notion of number in relation to fracvalue.  We introduce a way of specifying number systems, and compare the analytical concepts with those of structuralism.\\




\section{Introduction}

\begin{quote}
In the world of human thought generally, and in physical science particularly, the most important and fruitful concepts are those to which it is impossible to attach a well-defined meaning. 

Hendrik Anthony Kramers (1894-1952)\footnote{In M. Dresen, \textit{H. A. Kramers: Between Tradition and Revolution} (1987), 539.}
\end{quote}

\noindent The common meaning of `ambiguity', as expressed in the current (2020) \textit{Oxford English Dictionary}, is thus:
\medskip

`Originally and chiefly with reference to language: the fact or quality of having different possible meanings; capacity for being interpreted in more than one way; (also) lack of specificity or exactness.'
\medskip

\noindent The dictionary cites such usage from circa 1325. 

We will discuss ambiguity as it arises in a mathematical discourse where precision of expression is normally prized. Specifically, we will study ambiguity in the case of the word `fraction' in arithmetic.  In previous work \cite{Bergstra2020TM} 
on the extensive literature on (elementary, everyday) arithmetic, we have shown that `fraction'  \textit{is} ambiguous because it has a plurality of interpretations ranging from a syntactic construction---an expression with a division operator as its leading function symbol---to a number, this also qualifying as the value of a syntactic expression. Further, Fandino Pinilla~\cite{FandinoP2007} observes that fraction is not a mathmetical concept.

Therefore, here we examine the nature of its ambiguity, with the expectation of gaining some insights into the presence and role of ambiguous terminology in mathematics. For example, inevitably, the ambiguity of fraction leads us to consider the terminology `number'.

\subsection{The approach}
Consider a word $\mathsf{w}$ supposedly denoting a concept in some context. Now suppose 
that the word $\mathsf{w}$ is ambiguous, with a number of related interpretations in its context.\footnote{We are assuming a context in which the interpretations of $\mathsf{w}$ are related; words can be ambiguous with unrelated meanings, of course: such as the noun `nut'.}

Does ambiguity of the word $\mathsf{w}$ suggest that the intended concept is somehow vague or unclear? Even if it is primarily understood as the meaning of $\mathsf{w}$, close inspection reveals the meaning of $\mathsf{w}$ ramifies into different interpretations that are viable in the context; let us label them $\mathsf{concept_w}_1 \ldots, \mathsf{concept_w}_k$.

We will take the position that the ambiguity of $\mathsf{w}$ is a lexical matter to do with to the words of a language and what they might refer to:

\begin{assumption}\label{concepts_meaning}
Concepts are assumed to have unique meanings and ambiguities arise when several distinct concepts may be denoted by, or are bound to, a single word.
\end{assumption}

In the context of a mathematical discourse, Assumption \ref{concepts_meaning} reflects the working principle that concepts have precise definitions.
In some cases, recognising and embracing ambiguity is informative by experimenting with the following idea:

\begin{informal-proposal}\label{landmarks}
Suppose each of the interpretations labelled
$$\mathsf{concept_w}_1 \ldots, \mathsf{concept_w}_k$$
 is a `hypothetical meaning' of the term $\mathsf{w}$. Then, it is an option to make precise notions `near' these hypothetical meanings of $\mathsf{w}$ that allow clear definitions. We will call such clearly-defined `near notions' {\em landmark notions} for the term, this terminology reflecting a process or journey of exploring ambiguity. We will take the position that if it is hard or impossible to provide a satisfactory definition for the term $\mathsf{w}$ then what $\mathsf{w}$ refers to is not a concept.
\end{informal-proposal}

If the word is in common use then finding a definition of a concept without relying on the word $\mathsf{w}$ may turn out to be unpalatable, hard, if not impossible. In these circumstance the term need not be discarded.  From Brubaker \cite{Brubaker2025}, we take his use of the term \textit{category} for a bundle of related concepts all playing some role in a certain discourse. Thus,  the term $\mathsf{w}$ names a category containing the concepts labelled $\mathsf{concept_w}_1 \ldots, \mathsf{concept_w}_k$.

%
%
%

Assuming the ambiguity, then fraction will name a category while failing to be a concept (as it is ambiguous). We assume that concepts are categories as well, but not conversely: concepts have precise definitions, categories have illustrative descriptions. Below we will speak of fraction as a category and also as a word naming a category.

In analysing ambiguity we will use the more general term `notion' about which we have no specific views other than that both categories and concepts are notions. Notions may come with loose indications of theme, subject, meaning or relevance. Speaking of fraction as a notion abstracts from its status as a word. 

\subsection{Plan of the paper}
The paper has three phases. First, we will discuss ambiguities in general terms, including a general method of removing ambiguities (in Sections \ref{ambiguity_types} and \ref{removing_ambiguities}). Then we turn to fractions in the light of our general discussion (in Sections \ref{avoiding_fraction}, \ref{approximating_fraction} and \ref{frac_terminology}). The ambiguity of fraction is examined in considerable detail.

In the second phase, from the ambiguity of fractions we are led to reflect on the ambiguities of number. We approach number with a general method that distinguishes the lexical nature of ambiguity and the explicitness of particular constructions (in Section \ref{labels_shapes}).   

This necessarily draws us toward a large philosophical literature on number; here we tread lightly. In the final phase, we comment on ontology (in Section \ref{ontology}) and the established position of structualism (in Section \ref{structuralism}). With these in place we revisit the term fraction (in Section \ref{fractions_revisited}).
Given the ubiquity of division and therefore of fractions, in the light of our theorising, we address some `matters arising', including arithmetic at school level and some algebraic theories that motivate this paper, and observations on probability (in Section \ref{matters_arising}).


\section{Types of ambiguities in practice}\label{ambiguity_types} 

We begin with some observations on ambiguity.

\subsection{Controversial versus non-controversial ambiguities}

For all cases of lexical ambiguity it matters whether the ambiguity is controversial or not. In some cases that is easy to see. A clear example of a controversial ambiguity is with the use of the gender labels `male' and `female'. Some insist that a biological criterion is applied in determining male and female, and others claim that some form of apparently non-biological criterion are acceptable. Similarly, the nouns `man' and `woman' are ambiguous. Recalling the Informal Proposal  \ref{landmarks}, here landmarks are `biological male' and `person with male gender identity', and `biological female' and `person with female gender identity'.  We suggest that male and female are bi-ambiguous\footnote{Words are {\em bi-ambiguous} in a context when they have two meanings.} as their two landmarks are each seen as proper definitions by a large audiences. The ambiguity of gender labels (man and woman) has become remarkably controversial. 


In the context of mathematics, we consider the ambiguity of fraction to be {\em not} controversial. 


\subsection{Rigid ambiguity versus flexible ambiguities}

\begin{definition}
Ambiguity of a word is {\em flexible} if its various meanings or landmarks apply in different contexts and it is up to a user to disambiguate the word given its context. Ambiguity is {\em rigid} if it is not flexible, which means that users may choose certain landmarks for the word that they apply, or prefer to apply, in a context. 
\end{definition}

Perhaps polysemy is an adequate term for flexible ambiguity, but as far as we know a word is polysemous if, and only if, it is ambiguous.

\begin{definition}
A category is {\em rigidly ambiguous} if as a rule, in circles, groups, communities, cultures or localities it is unambiguous, while in different circles there may be different meanings. 
\end{definition}

A rigid category stands for a single concept in each circle, community, cultural locality in which it is used, it is ambiguous if it stands for different concepts in different circles or communities.

The gender labels man and woman are rigidly ambiguous concepts, because people tend to view these labels as unambiguous concepts, which is a shared meaning in their own circles or communities, irrespective of the fact that in other circles or communities people may take different concepts as the interpretation of these labels.   

\begin{claim}
Controversial ambiguity is likely to occur in case of rigid ambiguity.
\end{claim}

The ambiguity of man and woman is rigid as essentialists, who view gender as a synonym for biological sex, and co-essentialists, who view gender as a synonym for gender identity, will not compromise on their views in any setting.

The ambiguity of fraction is neither rigid nor controversial; it is locally rigid as in some circles or communities there may be well-defined concepts serving as the meaning of fraction.


\subsection{Significant ambiguity relative to a context}
Given that ambiguity cannot be moved out of the way entirely, what matters is the {\em significance} of the ambiguity of a word $\mathsf{w}$ relative to a context in which it is used. 

The gender labels man and woman feature significant ambiguity in many contexts (e.g., changing rooms). 

In the case of fractions, a listing of landmarks is considered in Section \ref{Landmarks_for_fraction} and shows the ambiguity to be significant technically: the significance lies in the observation that some landmarks allow certain features whereas other landmarks do not: ``having a numerator'', ``having a denominator'', ``being simplified'', ``admitting simplification''.  More explicitly, we may say the expression $\frac{4}{6}$ has an even denominator, while it is unconvincing to claim that the number $\frac{4}{6}$ has an even denominator. 



\subsection{Diffuse ambiguity with scattered landmarks}
As we have seen, not all manifestations of lexical ambiguity for potentially hypothetical word-bound concepts work the same way. A further distinction is necessary to explore fractions:

\begin{definition}
A word has {\em diffuse ambiguity} if none of its landmarks constitute a majority position, or even a significant minority position, among readers or users.  
\end{definition} 

We will argue that the ambiguity of fraction has these qualities: 
\medskip

- fraction features diffuse ambiguity with scattered landmarks

- fraction is flexibly ambigious, its meaning is context dependent

- fraction is not controversially ambiguous. 
\medskip

The presence of an ample plurality of landmarks for fraction will be illustrated in Section~\ref{Landmarks_for_fraction}. For a specific occurrence of fraction it is plausible that one of the landmarks provides a workable definition. In elementary arithmetic in the majority of cases it is possible to determine from the context of an occurrence of the word fraction which of the landmarks best captures the intended meaning of the occurrence.


\section{Removing ambiguities}\label{removing_ambiguities}

If a term is ambiguous then it has more than one interpretation or meaning that is sensible in a discourse. The question arises: 


\begin{question}\label{resolution_question}
Resolution: How can the meanings of a term $t$ that is ambiguous in a context $C$  be resolved in finitely many steps?
\end{question}

The term may have the virtue of being flexible, perhaps because the context of the term resolves any practical issues of meaning in some way; perhaps the meanings are rather close and their differences unlikely to have consequences. However, if there are different meanings then it might be of interest to examine them. Said differently, if a term has different meanings in a context then there must be different concepts at large in the context. With our Informal Proposal \ref{landmarks} in mind, the listing of possible meanings that are more precisely defined and their deployment raises the question:

\begin{question}\label{effects_question}
Effects: How is a discourse transformed if the meanings of a term $t$ that is ambiguous in a context $C$ are replaced by precisely defined landmark notions?
\end{question}

Here is our semiformal method for attempting to enumerate and classify these meanings.

\begin{method}\label{general_method}
Given a context $C$ and a term $t$ that has more than one meaning in $C$, we introduce new terms 
$$t_1, t_2,  \ldots, t_k,$$ 
each of which identifies a clearly definable interpretation of $t$ in $C$, i.e., a concept in $C$. Examining meanings for $t$ from the list creates a process of resolving its ambiguity. We consider the scope of $t$ to be defined by the list. We identify the priorities or biases in choosing meanings from the list, which may suggest an ordering of the interpretations determined by  $C$. If $t$ is influential in $C$ then we think of the $t_1, t_2,  \ldots, t_k,$ as {\em landmarks} in reading and understanding $t$. We think of $t$ as the name for a {\em category} of concepts.

Finally, we ask what are the consequences for a discourse about context $C$ of replacing the ambiguous term $t$ by some of the more precise terms from the list?
\end{method}

\section{Avoiding ``fraction''}\label{avoiding_fraction}
We will now apply this method to `fraction'.

\subsection{Why is fraction ambiguous?}\label{fraction_ambiguous}

To start, in arithmetic, the meaning of fraction is sometimes that of a structural notation, with a numerator and denominator, and sometimes is that of a number, often a rational number. For example, one may argue that all of the following are structural notations, but not all are numbers:
$$\frac{1}{2}, \  \frac{4}{8}, \  \frac{0}{2}, \  \frac{\frac{0}{2}}{2}, \ \frac{1}{0}, \ \frac{0}{0}, \ \frac{ \frac{1}{0}}{0}, \ \frac{ \frac{1}{2}}{ \frac{3}{4}}, \ \frac{ 0.5}{ 0.75}.$$
Indeed, fractions  involve operations in their numerators and denominators, e.g., as when one simplifies $\frac{ \frac{1}{2}}{ \frac{3}{4}}$ to $\frac{1\cdot 4}{2\cdot 3}$ and $\frac{4}{6}$. Actually, teaching fractions in school---a core activity in basic arithmetic and the subject of a vast literature---is teaching a semiformal calculus for combining and transforming fractions. 

The origin of this analysis is our observation---or perhaps phrased more cautiously, our proposition--- that the notion of a fraction \textit{defeats} being given a convincing definition which covers the various uses of fraction in elementary arithmetic; and, in contradiction to Kramers's observation, that this is a problem. For a justification of the proposition on the lack of a convincing definition of the concept of fraction we refer to~\cite{Bergstra2020TM}.\footnote{This belongs to a  line of technical work including~\cite{Bergstra2020SACS,BergstraP2016IM,BergstraP2020AI,BergstraT2023a,BergstraT2024b}.} 

Attempts to find a convincing definition fail because the ambiguity is intrinsic: the underlying idea somehow combines syntax and semantics into an amalgam which can play different roles. The taxonomy of fractions, like \textit{simple fraction}, \textit{unit fraction}, \textit{simplified fraction}, \textit{composite fraction}, $etc.,$ 
uses structure and suggest a syntactic interpretation of fraction and are possible landmarks in the sense of Informal Proposal \ref{landmarks}.
Similarly, the idea that a fraction is a `part of a whole' suggests a semantic bias.


Instead of looking for a convincing definition of fraction we look for notions that can be more easily and precisely defined, and that might form a basis of explanations of elementary arithmetic involving division. We follow the method and introduce several new terms to distinguish some of the meanings associated with fraction. Such notions will be termed landmarks  in the context of a search for the meaning of fraction. Specifically, we introduce the key notions 
\begin{center}
\textit{fracterm, fracvalue, fracsign} and \textit{fracsign occurrence}. 
\end{center}
Our basic idea is to use the term \textit{fracterm} to make precise the structural interpretation of fraction and \textit{fracvalue} for the numerical interpretation. In this case, the four notions specify different levels of abstraction in the presentation of certain mathematical entities and become our main landmark notions, which are more precisely and easily explained than the target category of fraction. 
Actually, with these aspects precisely defined, we can make the ambiguity of fraction explicit, by synthesising the notion and term of a \textit{fraxion} as the disjoint union of the four notions just mentioned.  
Here are the four new concepts.


\subsection{Four concepts and terms}
Fracterm is a notion which lies at this syntactic side of the range of mathematical concepts; its name is a shorthand for `fractional expression'.  The separation between syntax and semantics is not sharp in mathematics unlike in logic. Notions like power series and matrix lie somewhere between syntax and semantics. 
Fracterm shares its syntactic attribute with commonplace mathematical notions \textit{polynomial}, \textit{quadratic form} and \textit{normal form}. Whilst in logic mention of `term' is indicative of a role for syntax, in mathematics occurrences of `term' are less usual while the occurrence of `form' is indicative of a similar role. 


\subsubsection{Fracterm: a syntactic notion}
A fracterm is a syntactic expression with division as its leading function symbol. Formally, a signature $\Sigma$ is a list of the names of the basic constants and operators from which a set of algebraic expressions are built. To have a more precisely focused notion:

\begin{definition}\label{def: fracterm}
Let $\Sigma$ be a signature containing a division operator.  A $\Sigma$-fracterm is a $\Sigma$-term with division as its leading operator symbol. 
\end{definition}

For definiteness, we will assume that our terms are based on a signature with constants $0, 1$ and operations $+, -, \cdot, \div$.
\footnote{Occasionally, other symbols such as $\bot$ will be added as a constant.} 

The examples in the list in section \ref{fraction_ambiguous} are all fracterms, as are more wayward recursive examples involving $\frac{1}{0}$ and $\frac{0}{0}$ and variables $x, y$, such as 
$$\frac{\frac{1}{0}}{\frac{0}{0}}, \ \frac{\frac{0}{0}}{\frac{0}{0}}, \frac{\frac{x}{0}}{\frac{0}{0}}, \ \frac{\frac{x}{0}}{\frac{y}{0}}, \ldots$$

In proposing to use fracterm, we give ourselves a sharply focussed basic notion. By its syntactic nature, it is more confined than the common conventional understanding  of fraction as it excludes quantification via numbers, ratios, or references to parts of a whole.  Yet, it is less confined in  
embracing \textit{all} arithmetic expressions with division as the leading function symbol. In particular, it avoids premature limitations to: closed expressions; integer or natural number numerators and denominators; non-zero denominators;  division-free numerators and/or denominators;  each being restrictions which often come with explanations of fractions. 

Applying the taxonomy of fractions, there are several familiar ways to qualify the fracterms:

\begin{definition}\label{flattening}
A {\em flat fracterm} is a fracterm of which the numerator and denominator do not involve division.
\end{definition}

 A fracterm may allow simplification to a flat fracterm: for instance a transformation of the form
$$ \frac{\frac{a}{b}}{\frac{c}{d}} => \frac{a \cdot d}{b \cdot c}.$$ 
We notice that flatness is {\em not} a property of fracvalues.


\begin{definition}\label{simple_fracterms}
A {\em simple fracterm} is a flat fracterm that has (decimal) integer numerals as its numerator and denominator.  A {\em simplified simple fracterm} is a simple fracterm which does not allow further simplification. The simplified simple fracterms we denote $SSFT$.
\end{definition}

Simplified simple fracterms are syntactic entities close to common representations of rational numbers.

\subsubsection{Fracvalue: a semantic counterpart for fracterm}\label{fracvalues}

\begin{definition}
The syntactic notion of fracterm is intended to denote a {\em fracvalue}. Fracvalues are typically numbers but in practice also include  elements such as 
$$ \bot, \infty, +\infty, -\infty, NaN$$
which we call {\em peripheral numbers}.\footnote{NaN is a common peripheral in computer arithmetic, standing for 'not a number'.} We assume that all fracterms denote fracvalues. 
\end{definition}

In arithmetic, fracterms denote numbers, a seemingly obvious observation, but defining what these numbers are is not an easy matter at all. For example, different constructions, with and without peripherals, for rational numbers provide different collections of fracvalues. Thus, in the context of fields with division that are common meadows, the fracterm $\frac{1}{0}$ has the fracvalue and peripheral number $\bot$, \cite{BergstraT2023}.


\subsubsection{Fracsigns}
The terms fracterm and fracvalue are mathematical concepts. At the other end of the spectrum of meanings that we are exploring there is the textuality of fracterms and fracvalues. Our terminology for this aspect we call {\em fracsigns}.

Fracsigns may refer to entities  other  than fracterms or fracvalues. Any text or context may steer the way how signs are to be understood. So, the rule that a fracsign refers to a fracterm, or to a fracvalue, may be understood as a {\em default} rule, which by its definition allows for exceptions.

Fracsigns in turn come in formats. For instance, for division there are these operator formats
$$\frac{\star}{\star}, \ \  \star \div \star, \ \  \star / \star, \ \  \star\, \colon \star $$
We will often use the inline format $\star \div \star$, and we will not attend to the variety of operator formats, or to the rationale of using different formats, which is a theme in its own right. 

The meaning of a fracsign $t \div r$, where $t$ and $r$ are terms, is open-ended and depends on the context of use. It is reasonable to view fracsigns as basic entities in elementary arithmetic, which serve as one of the points of departure for the explanation and practice of rational arithmetic. 
That fracsigns are symbolic means they can be further decorated to make explicit their levels: for instance, $t \div_{\mathsf{ft}} r$ where $t \div r $ is understood as a fracterm; and 
$t \div_{\mathsf{fv}} r$ where $t \div r $ is understood as a fracvalue.  These decorated fracsigns leave less space for level assignment and ambiguity than undecorated fracsigns. Alternatively, one may view the subscript as a feature of context.


\subsubsection{Fracsign occurrences}

A page of text may contain different occurrences of the same fracsign, such occurrences may further show variation in typography, size, colour etc. The lowest level of abstraction, or level for short,  we consider is the {\em occurrence of a fracsign}. As elements of some level, it is occurrences---rather than fracsigns--- that are embedded in a context and that may impact the level assignment of the fracsign occurrence at hand.


\subsubsection{On terminologies}
While we consider the word `fracterm' to be clear and convincing, the situation with `fracvalue' may seem less so---why not just `number'? 
Using fracvalue is only plausible in combination with fracterm. As has been amply demonstrated in computer science, where syntax is a primary object of study and semantics is its necessary but complex companion, syntax increases the precision and rigour of an analysis, and in the case of fractions the scope of an analysis (e.g., accommodating $\frac{1}{0}$ and peripherals).

One can introduce a special term: a fracvalue that is not a peripheral element might be called a {\em fracnum}.
While most numbers are fracnums there is hardly a significant role for a notion of a fracnum besides the notion of a number.

For the role of fracsign we have also considered as alternatives `fracsymbol' and `fractoken'. We have chosen fracsign because symbol has an `atomic' connotation as a primitive entity, which sign has not (or in any case less), while fractoken suggests a mere (physical) entity which represents another entity, which is not the case for sign.

Ratio is a word similar to fraction, with uses inside and outside arithmetic, though more informal. When used in isolation, ratio tends towards a pair such as a {\em staff:student ratio} $x:y$ or some gambling odds. However, when used as component of a phrase it may be otherwise: for instance, the statistical notion(s) of {\em likelihood ratio} is a rational number, and the {\em payout ratio} of a company will be a rational number.
At this stage we will assume that fraction and ratio are ambiguous in the same way, so that ratio also features diffuse ambiguity.


\subsection{Landmarks for fraction}
\label{Landmarks_for_fraction}
Below we list some landmarks for fraction, which are related in several ways, most being self-evident.
\begin{enumerate}
    \item fracterm, that is an expression of the form $\frac{P}{Q}$;
    \item fracsign occurrence, a sign (e.g., on paper or a blackboard, or a screen) denoting a fracterm;
    \item fracsign, the abstraction of all fracsign occurrences;
    \item  flat fracterm, fracterm $\frac{P}{Q}$ where the expressions $P$ and $Q$ do not     contain occurrences of division;
    \item simple fracterm, a fracterm $\frac{P}{Q}$ where $P$ and $Q$ are decimal integers, 
    \item safe simple fracterm, a fracterm $\frac{P}{Q}$ where $P$ and $Q$ are decimal integers, and $Q$ is nonzero (for instance $\frac{17}{-332}$);
    \item \label{SSSFT} simplified safe simple fracterm (for instance al integers, and $Q$ is nonzero (for instance $\frac{-3}{7}$ but not $\frac{-3}{-9}$);
    \item \label{Fracpair} fracpair, a pair of integers;
    \item safe fracpair, a pair of integers, with the second one non-zero;
    
    \item a fracvalue (named)  by a fracterm (not a mathematical notion, with naming a mechanism outside mathematics, this idea has several variations by restricting names to various subclasses of fracterms);
    \item fracvalue, a value of a fracterm, plausibly a rational number (close to: a fraction is a part of an integer number);
    \item a quotient of two integers, the second one nonzero;
    \item a rational number which is not an integer number;
    \item fraxion: a fraxion is either a fracsign, or a fracsign occurrence, or a fracvalue (fraxion is one of a plurality of possible disjunctive interpretations of ``fraction'', see also~\ref{Fraxion} below).
\end{enumerate}

Notice that ambiguity persists: it is not the case that the various landmark notions leave {\em no} doubts about precise meaning. For instance, with fracterm, one may ask in which signature the terms are written. The ambiguity of fraction matters in elementary arithmetic because some key notions are sensitive for said ambiguity. Thus, landmarks are meant to be less ambiguous, and so ambiguity is likely to resurface at a smaller scale.


\subsection{Levels of abstraction}\label{basic_abstraction}
Fracsign occurrence, fracsign, fracterm, and fracvalue are four levels of abstraction that underly the concept of a fraction. The basic `hierarchy' is, from low to high levels,
\begin{center}
fracsign occurrence $\to$ fracsign $\to$ fracterm $\to$ fracvalue. 
\end{center}
Actually, there are abstraction levels in between, e.g., assuming only closed expressions one may position simple fracterm and  simplified simple fracterm between fracterm and fracvalue (Definition \ref{simple_fracterms}). Per abstraction level, mathematical definitions are possible, though giving such definitions is not without issues, because when proposing a definition a choice must be made from a range of options. The definition of addition for simple fracterms may be given in several ways, which may deliver different fracterms for the same fracvalues,  e.g., addition when applied to $\frac{1}{2}+ \frac{3}{2}$ may produce $\frac{4}{2}$, or 
$\frac{8}{4}$, or $\frac{2+6}{4}$.

Now, the notion of `fraction' resists being identified as belonging exclusively to any one of these particular abstraction levels. Thus, the status of the term `fraction', if it is not redundant, may be taken to be an intuition that underlies and motivates its various projections on to the more clearly definable landmarks. To make this ideas clear we define: 

\begin{definition}\label{def:cross-cutting}
By a {\em cross-cutting category} we mean a category that captures and denotes a common identity or purpose for distinct concepts but rejects the need for its precise analysis and formalisation.
\end{definition}

To test these notions, first we will make an attempt to approximate a mathematical definition of fraction using them: we will not succeed. Although we expect that in fact {\em no} satisfactory definition of fraction can be given, we accept that cross-cutting notions such as fraction have a role and, indeed, cannot be avoided in practice when working with elementary arithmetic.


\section{Approximating `Fraction'}\label{approximating_fraction}

Having applied the method to answer the Resolution Question \ref{resolution_question}, we have precise terms for conceptual ingredients of fraction; can we synthesise a precise interpretation of the ambiguous notion?


\subsection{Attempt 1: Fraxion} 
\label{Fraxion}

We introduce the term \textit{fraxion} for an entity which is either a fracvalue, fracterm, or a fracsign; in symbols, it is a disjoint union of concepts,
\begin{center}
fraxion = fracvalue $\oplus$ fracterm $\oplus$ fracsign  
\end{center}
Fraxion is an unambiguous category and, since it is precisely defined, a concept.

Fraxion is designed to  leave open the choice between the three related levels. Rather than being an ambiguous category, fraxion is a disjunctive concept, a concept designed to allow for a definite degree of semantic freedom. An ambiguous category has no definition in which choice of meaning is part of the definition, whereas a disjunctive concept, has choice explicitly built in as a 
part of the definition. In using fraxion, it needs to be assumed that which of the three options applies can be inferred from the context. For a text there may be rules governing level inference for fraxions. For instance, it may be stated that a fracsign occurrence is understood as a fraxion and that in case of the absence of other information a fraxion is taken for a fracvalue. 

For certain purposes, an occurrence of fraxion may be disambiguated by tags: fraxion$_{\mathsf{ft}}$, fraxion$_{\mathsf{fv}}$, fraxion$_{\mathsf{fs}}$ and fraxion$_{\mathsf{fso}}$.

In fraxion we incorporate precise descriptions of different forms of ambiguity. It is the ambiguity represented by the component 
fracvalue $\oplus$ fracterm that is important for arithmetic in the case of division. We arrive at two claims which limit the clarity and appeal of any definition of fraction:

\begin{claim}
A defnition of fraction---if any definition can be provided---cannot be expected to resolve ambiguities which are intrinsic to natural language, inside or outside arithmetic.
\end{claim}
\begin{claim}
A definition of fraction---if any definition can be provided---cannot be expected to be more clear and less ambiguous than any definition of natural number. 
\end{claim}

Fraction and fraxion are {\em} not the same notions, but the notions share properties. Fraxion is a word which takes different meanings in different contexts. It is not the case, however, that the concept fraxion can be understood as the totality of all meanings in all possible contexts, simply because it is the union of fracterms, fracvalues, and fracsigns.

We find that fraxion is a word which refers and refines to  fracterm,  fracvalue, or  fracsign, where each option for refinement can be taken only if the context of use of fraxion admits that option. In a text, in most cases occurrences of fraxion may thus be replaced. At the same time `a fraxion' is any entity which fraxion may refer to (i.e., an element of the union of the kinds in the disjunction).


\subsection{Attempt  2: Approximating the use of `fraction'}

One may use fraxion to refer to any of its component notions. Can we go further in pursuit of a precise notion that approximates the category fraction more closely, as it is used in the context of arithmetic? 

\begin{claim}\label{approx_fraction}
Consider a notion which is equipped with a {\em bias} when referencing the three components that make up fraxion.
The notion works like fraxion, in that each occurrence must be refined into one of fracvalue, fracterm, and fracsign. However, this refinement is to be done with these priorities: 

1. if refinement to {\em fracvalue} fits in the context then that choice must be made; else 

2. if refinement to {\em fracterm} fits in the context then that choice must be made; else

3. refinement into {\em fracsign} is the only remaining option and that choice must be made.
\end{claim}

Adding these constraints to fraxion creates another disjunctive level, though one less ``ambiguous'' than fraxion. For this notion we {\em might} be tempted to introduce a new term.

These constraints provide a workable interpretation for fraction because in practice:

(i) fraction is an ambiguous level,

(ii) fraction allows the same refinements,

(iii) fraction involves the same preferences for refinement.

Thus, in this way, we can identify aspects of the use of the term fraction thanks to these four notions. To invent a new precise term however is not appropriate. Perhaps, following Kramers as cited in the introduction, the use of fraction is to flag the existence of, and to generate, new precise concepts.

Although we can imagine refinements, we will propose fraxion as a well-defined concept which is closer to `fraction' than either fracterm or fracvalue or fracsign. 

However, for analytical purposes, we find no advantage in {\em using} fraxion in addition to fracterm and fracvalue, so that fraxion turns out to be merely our best shot at explaining what fraction might mean. We have experimented with the constraints of Claim \ref{approx_fraction} to make a better approximation of `fraction', but we were unable to clarify the ambiguous aspects of `fraction' any better than `fraxion' does. 

However, a proposal for less ambiguous language can be made: instead of using fraction,  

(1) use fracterm rather than fraction in accounts of arithmetic whenever a syntactic bias were plausible.  

(2) use fracvalue rather than fraction (or a semantic notion like rational number or quotient), whenever a semantic bias were plausible.


\section{On `frac' termininology in practice}\label{frac_terminology}

To complete the application of the method, we must try to answer the more expansive Effects Question \ref{effects_question}. 

We now consider texts  involving our terminology of fracsigns, fracterms, fracvalues and fraxions, which we call \textit{fractalk}. On meeting a fracsign occurrence in a text any level of interpretation, or a combination of such levels levels may arise. Using simple texts as examples of contexts, we will explore how our new terminology performs.

The texts are sequences of simple assertions. Texts are read in the order of presentation of these assertions. An assertion may refer to occurrences of fracsigns anywhere in the sequence, viz. in earlier or later assertions.  Each assertion is claimed to be true.

For fracvalues, we assume a presentation of the rational numbers with the property that no number is also fracterm at the same time. 


\subsection{Fractalk I: The need of a distinction between fracterms and fracvalues}
\label{Fractalk1}
We give an invalid argument due to ambiguity.
\begin{itemize}
\item {\em Assertion sequence A.}

The first assertion sequence demonstrates the argument for making a distinction between fracterms and fracsigns. Here fraxion is used for ``one's preferred interpretation of a fracsign occurrence''. Consider the assertion sequences which are composed of the following five assertions:
\begin{enumerate}
\item the fraxion $\frac{1}{2}$ has numerator $1$;
\item the fraxion $\frac{2}{4}$ has numerator $2$;
\item fraxions have a unique numerator;
\item $\frac{1}{2}$ equals $\frac{2}{4}$;
\item combining the above assertions:  $1= 2$.
\end{enumerate}
This argument may be formalized using an operator $\mathsf{Num}(-)$ which extracts the numerator from a fracsign, and may be introduced in combination with an operator $\mathsf{Denom}(-)$, which extracts the denominator from a fracsign.
 $$1 = \mathsf{Num}(\frac{1}{2}) = \mathsf{Num}(\frac{2}{4})= 2$$
 Here, we assume that the operations $ \mathsf{Num}$ and $ \mathsf{Denom}$ satisfy for all fracterms, fracsigns, and fracsign occurrences $P$, the following requirement:
 $$P = \frac{\mathsf{Num}(P)}{\mathsf{Denom}(P) }$$
 
 We consider the derivation of $1=2$, which may be understood as a self-contradictory claim, to be so problematic that a preferred argument against its validity needs to be incorporated in any satisfactory account of elementary arithmetic. Different arguments against the argument in assertion sequence A may be brought forward, and our preference is to consider {\em confusion of levels} as the source of the problem, leading to an account of the situation as given in the following assertion sequence A$'$.
 
 The paradoxical reasoning leading to $1=2$, as demonstrated above, 
 and its various weaknesses were discussed in extensive detail in~\cite{Bergstra2020TM}.
 
\item  {\em Assertion sequence A$'$}.

Using fracterms and fracvalues for disambiguating occurrences of fraxion the assertion sequence A may be rewritten in such a manner that the contradiction now disappears (while turning into an obviously invalid argument) due to a distinction of the levels fracterm and fracvalue (where we once more assume that fracvalues are chosen from a shape for rationals in which no number is a fracterm at the same time):

\begin{enumerate}
\item the fracterm $\frac{1}{2}$ has numerator $1$;
\item the fracterm $\frac{2}{4}$ has numerator $2$;
\item \label{three} fracterms have a unique numerator;
\item \label{four} the fracvalue $\frac{1}{2}$ equals the fracvalue $\frac{2}{4}$;
\item combining the above assertions:  $1= 2$. 
\end{enumerate}
The final inference has become invalid because 
assertion~\ref{four} is about fracvalues and the relevant uniqueness has only been established in assertion~\ref{three} about  fracterms, and not about fracvalues. We find that there is no paradox left, merely an assertion sequence ending in a mistaken conclusion which comes about from discarding the distinction between fracterms and fracvalues.
\end{itemize}


\subsection{Fractalk II: The need for a level fracsign besides fracterm and fracvalue}
It seems that by introducing a distinction between the levels fracterm and fracvalue, the ambiguity underlying the paradox suggested in assertion sequence $A$ has been removed, thereby obtaining a fully waterproof understanding of fractalk. That suggestion, however is too optimistic. Upon distinguishing fracterms and fracvalues another issue arises, as is documented in the following assertion sequence. 

We assume that numbers are chosen in such a manner that fracterms and fracvalues are disjoint collections. 
\begin{itemize}
\item  {\em Assertion sequence B}.
\begin{enumerate}
\item $\frac{2}{3}$ is a rational number;
\item $\frac{2}{3}$ is a fracterm;
\item \label{three2} rational numbers are not fracterms;
\item $\frac{2}{3}$ contradicts the claim in item~\ref{three2}.
\end{enumerate}

These assertions together indicate a fundamental ambiguity of the fracsign $\frac{2}{3}$: it may refer to a fracterm and it may refer to a fracvalue and these have just been determined/assumed to be disjoint (as a response to the issue raised in assertion sequence A).

We find that the sign $\frac{2}{3}$ as it occurs in three lines of assertion sequence B has different interpretations. In assertion 1 it is a fracvalue, in assertion 2 it is a fracterm and in assertion 4 its interpretation is undecided. We find that the sign (fracsign) $\frac{2}{3}$ is ambiguous and that its various presences (= fracsign occurrences) have different interpretations depending on the context. The example provides ample justification of considering both fracsign and fracsign occurrence as levels of abstraction of relevance for rational arithmetic.

\item  {\em Assertion sequence B$'$}.
The assumption that fracterms and fracvalues are disjoint is inessential for the above example.
It is possible to choose number collections in such a manner that all rational numbers are fracterms, for instance by using the $SSFT$ definition for rationals (Definition \ref{simple_fracterms}).  In the $SSFT$ definition the fracterm $\frac{4}{6}$ is not a rational number, while its simplification  $\frac{2}{3}$ is a rational number so that assertion 3 of assertion sequence B would be wrong.

Nevertheless, it is plausible to claim that ``$\frac{4}{6}$ is a rational number'' as a judgment about the fracsign occurrence at hand, or as a judgment about the corresponding fraxion. We find assertion sequence B$'$ which parallels assertion sequence B
\begin{enumerate}
\item $\frac{4}{6}$ is a rational number;
\item $\frac{4}{6}$ is a fracterm;
\item \label{three2}  not all fracterms are rational numbers, in particular not the fracterm $\frac{4}{6}$;
\item $\frac{4}{6}$ contradicts the claim in item~\ref{three2}.
\end{enumerate}
Now these assertions together indicate an ambiguity of the fracsign $\frac{4}{6}$: it may refer to a fracterm and it may refer to a fracvalue and these levels have been shown different in certain cases, even if some fracterms are fracvalues at the same time. Again we find that thinking in terms of different fracsign occurrences for the same fracsign (three such occurrences in assertion sequence B$'$) makes much sense. In particular, one reads assertion 1 as a claim about the fracsign occurrence in that line, while assertion 2 is read as a claim about a different fracsign occurrence and combining both claims as is done in 4 is mistaken.

\item  {\em Assertion sequence B$''$}.

We rewrite assertion  sequence B once more just by commenting its parts. The idea is that disambiguation of fraxion depends on the occurrence of the fracsign at hand so that different occurrences (at least occurrences in different assertions in the sequence) of the same fracsign may be given different levels. Unlike in assertion sequence B$'$ we will now assume, as in assertion sequence B, that fracterms and fracvalues are disjoint levels.
\begin{enumerate}
\item $\frac{2}{3}$ is a rational number; (in this line the fracsign occurrence is understood as a fraxion which is understood as a fracvalue)
\item $\frac{2}{3}$ is a fracterm; (in this line the fracsign occurrence is understood as a fraxion which is understood as a fracterm)
\item \label{three3} rational numbers are not fracterms;
\item $\frac{2}{3}$ contradicts the claim in item~\ref{three2}.
\end{enumerate}

Assertion 4 is now clearly false and not somehow paradoxical because for neither interpretation of $\frac{2}{3}$ as a fraxion it is the case that it is both a fracterm and a fracvalue at the same time.
\end{itemize}


\subsection{Failed paradox reconstruction with forward referencing}
As we did for assertion sequence B we will adopt the assumption that fracterms and fracvalues are disjoint collections.
We understand a paradox as the occurrence of a false assertion in an assertion sequence in which each assertion is either obvious from first principles or follows or seems to follow from earlier assertions in the same assertion sequence. We may try to reconstruct the paradox as alluded in assertion sequence B by means of forward references to fracsign occurrences. For instance:
\begin{itemize}
\item {\em  Assertion sequence C.}
\begin{enumerate}
\item \label{one} the occurrence of the unique fracsign in the following assertion has level fracterm;
\item \label{two0} $\frac{2}{3}$ is a rational number;
\item $\frac{2}{3}$ is a fracterm and a fracvalue at the same time;
\item \label{three4} rational numbers are not fracterms;
\item $\frac{2}{3}$ contradicts the claim in item~\ref{three4}.
\end{enumerate}
Here, however, assertion~\ref{two0} may be considered clearly false on the basis of assertion~\ref{one}, so that no paradox arises.
The situation is made more explicit in assertion sequence C$'$ below.
\item {\em Assertion sequence C$'$}.

\begin{enumerate}
\item the occurrence of the unique fracsign in the following assertion has level fracterm; (reading from left to right in order to make this assertion true it will be assumed that the fracsign in assertion~\ref{two} refers to a fraxion which is taken for a fracterm.)
\item  \label{two}  $\frac{2}{3}$ is a rational number; (now this assertion has become false, as the option to assign the occurrence of   $\frac{2}{3}$ the level fracvalue has been ruled out in assertion 1.)
\item $\frac{2}{3}$ is a fracterm and a fracvalue at the same time; (this claim does not hold either anymore.)
\item \label{three5} rational numbers are not fracterms (true by assumption);
\item $\frac{2}{3}$ contradicts the claim in item~\ref{three5}; (the contradiction has been removed because of invalidity of assertion~\ref{two} in the sequence).
\end{enumerate}
\item {\em  Assertion sequence D}.

Working with fraxions a paradox seems not to arise, but the language will be too weak for practical fractalk.

\begin{enumerate}
\item $\frac{2}{3}$ is a fraxion;
\item all rational numbers are fraxions;
\item not all fraxions are rational numbers;
\item $\frac{2}{3}$ may be a rational number; (or more precisely: an occurrence of a fracsign denoting the fraxion $\frac{2}{3}$ may sometimes be read as a fracvalue, but it need not be read as such.)
\end{enumerate}
\end{itemize}


\subsection{Fractalk III: Referring}
The following rule captures how references work in elementary arithmetic:
\smallskip

\noindent {\it Reference chaining by default rule}: If P is a reference to Q and Q is a reference to R then P is understood as a reference to R unless the level ``reference to Q'' is made explicit. 

Following this rule, a fracsign occurrence is usually supposed to refer to a most abstract referent given the context, which in most cases is a fracvalue.
\smallskip

As examples of this rule we mention: 
\begin{itemize}
\item ``we know that $\frac{4}{3} >1$'' -- so, $\frac{4}{3}$ defaults to fracvalue
\item ``$\frac{4}{3}$ is both simple and simplified'' -- so, $\frac{4}{3}$ defaults to fracterm
\item ``$\frac{4}{3}$ measures the relative size of the part of the cake that has colour blue'' -- so, $\frac{4}{3}$ defaults to a fracvalue
\item in ``$\frac{4}{3}$ is  improper'' -- so$\frac{4}{3}$ defaults to fracterm
\item ``the fracterm $\frac{4}{3}$ is both simple and simplified''
\item In ``Is the sign $\frac{4}{3}$ in the following equation a reference to a fracterm or to a value: $\frac{4}{3} = 1 + \frac{1}{3}$?'', the first occurrence of $\frac{4}{3}$ defaults to fracsign. (Where we notice as a side remark that the question has no definite answer, both interpretations are possible.)
\end{itemize}
It is highly unusual in elementary arithmetic that a metavariable for fracterms is used. Although a text fragment like ``let $P$ be the fracterm $\frac{2}{5}$ ...'' is perfectly comprehensible, it is also rather exotic. This convention limits in practice  the options for application of the chaining by default rule.

In the context of arithmetic with division, in line with  the hierarchy, the following chain of references exists: 
\begin{center}
fracsign occurrence $\to$ fracsign $\to$ fracterm $\to$ fracvalue. 
\end{center}
Using the chaining rule a reference made by an occurrence of a fracsign is first of all understood as a reference to the fracvalue (i.e., to the final stage in the chain of references). Only the presence of unambiguous level information may shorten the chain of references. For instance ``as a result we find the fracterm $\frac{p}{q}$ for which we may compute an approximate value.''

 
 
  
 
 


\subsection{Fractalk IV: Everyday assertions involving fraction}
\label{Fractalk4}

Given the notions of fracsign occurrence, fracsign, fracterm, fracvalue, fraxion and some rules for their deployment, are there rules that guide the unambiguous use of fraction in everyday arithmetical talk? 

We have collected some such rules for reading/interpreting occurrences of ``fraction'' in fractalk. The idea is that in practice nearly each occurrence of fraction is in some need of disambiguation, and that mainly fracterm (or qualified fracterms such as simple fracterm, unit fracterm etc.) and fracvalue (rational number, quotient) can be used for that purpose while the need for using fracsign occurrence, fracsign or fraxion will be much less frequent. In some cases no disambiguation is in order, for instance in the following question: ``What is a fraction?''. 

If one encounters say, the following claim ``a fraction is a rational number'' disambiguation is not a merely local matter. Locally, that is for the occurrence of the word fraction at hand, reading it as fracvalue works. But the claim suggests that in the remainder of the text (in which the claim occurs) fraction will stand for fracvalue and not for fracterm or fracsign or fracsign occurrence or any other landmark for fraction. This claim is supposedly meant to be helpful for reading the text in which it is embedded, while it does not constitute a definition. 

A claim which supposedly contributes to understanding the meaning of a phrase, though which may fall short of constituting a definition, will be referred to as a \textit{definitional claim}. 

Consider the following assertion sequence $E$:
\begin{itemize}
    \item {\em Assertion sequence E.} 
\begin{enumerate}
    \item Definitional claim: a fraction is a number;
    \item The fraction $2/6$ has an even denominator.
\end{enumerate}
\end{itemize}
Disambiguation of assertion sequence $E$ is problematic because we do not assign a denominator to the value $2/6$ (an extensive analysis of this issue can be found in~\cite{BergstraP2020AI}). 
If hard pressed  one must determine or suggest a denominator of a number nevertheless, the plausible idea is to view the value $2/6$  as the value of the unique simplified simple fracterm $1/3$ and to assign as a denominator $3$ to it, which is at odds with ``the denominator of the fracsign $2/6$''. 
In case taking a definitional claim too literally creates a problem with subsequent disambiguation, such a definitional claim   may probably best be disambiguated so as to become trivial, for instance by casting ``a fraction is a number'' as a ``a fracvalue is a number'', thereby avoiding subsequent level inference difficulties.
\begin{itemize}

\item  {\em Assertion sequence F.}

Each use of ``fraction'' which can be easily disambiguated (into fracsign occurrence, fracsign, fracterm, or fracvalue) is legitimate and can be used accordingly. Accordingly the following assertions are unproblematic, recalling definitions earlier (in \ref{basic_abstraction}):

- the fraction $4/2$ is an even integer;

- the fraction $2/4$ can be simplified;

- the fraction $1/2$ equals the fraction $2/4$;

- the fraction $\frac{1+ 2/3}{5}$ is not flat;

- $5/(1+3)$ is not a simple fraction;

- $5/(1+3)$ is not a simplified fraction (although it cannot be simplified);

- $5/(1+3) $ can be written as a flat fraction (for instance as $5/4$);

- $5/4$ is not a proper fraction;

- $\frac{1+ 2/3}{5}$ can be written as a flat fraction (for instance as $5/15$).

Disambiguation rules for fraction are best designed in such a manner that as many texts as possible are given a proper interpretation.

\item  {\em Assertion sequence G.}

\begin{enumerate}
\item Equivalence of simple fracterms $\frac{a}{b}$ and $\frac{c}{d}$, written 
$\frac{a}{b}=\frac{c}{d}$ is defined by: ($b = 0$ and $c = 0$) 
or ($b\neq 0$ and $c \neq 0$ and $ a \cdot d = b \cdot c$).

\item Addition is a family of operations consisting of several related components:
\begin{enumerate}
\label{addFamOp}
\item
 a function on (rational) numbers, and at the same time (by way of overloading),

\item
\label{nonDetAdd} a nondeterministic function (i.e. relation) on flat fracterms, and (once more by way of overloading),

\item
\label{nonDetAdd2} a nondeterministic function (i.e. relation) on simple fracterms, and (once more by way of overloading),

\item 
\label{nonDetAdd3} a nondeterministic function (i.e. relation) on simplified simple fracterms, and (once more by way of overloading),

\item a nondeterministic function on arithmetical expressions.
\end{enumerate}
In each case we may write $+_{\mathsf{add}}$ for said (nondeterministic) function
\item For instance (w.r.t.~\ref{addFamOp} no.~\ref{nonDetAdd}) on flat fracterms addition may (but need not) be achieved as follows, with $a,b,c,d$ division free expressions:
$$\frac{a}{b} +_{\mathsf{add}}\frac{c}{d} = \frac{a\cdot d + b \cdot c}{c \cdot d}.$$
Alternately (that is an alternative way of addition incorporated by $+_{\mathsf{add}}$) may separate the case that $b$ and $d$ are identical expressions in which case a simpler determination of the result can be given:
$$\frac{a}{b} +_{\mathsf{add'}}\frac{c}{b} = \frac{a+ c}{b}$$ and in all  other cases
$$\frac{a}{b} +_{\mathsf{add'}}\frac{c}{d} = \frac{a\cdot d + b \cdot c}{c \cdot d}$$

\item Now w.r.t.~\ref{nonDetAdd2}: on simple fracterms addition may (but need not) work as follows, with $a,b,c,d,e,f$ decimal integers:
$$\frac{e = a\cdot d + b \cdot c, \, f = c \cdot d}{\frac{a}{b} +_{\mathsf{add}}\frac{c}{d} = \frac{e}{f}}.$$

\item And w.r.t.~\ref{nonDetAdd3}, on arithmetical expressions addition may (but need not) trivially work as follows, with $P$ and $Q$ arithmetical expressions which are not fracterms:
$$P +_{\mathsf{add}}Q  =P + Q.$$
\end{enumerate}

\end{itemize}

\section{Numbers as labels and shapes}\label{labels_shapes}

The coherence and stability of theories of syntax are not complemented by those of theories of semantics, which are many, varied and not always comparable or consistent. This is vivid in the development of programming languages since the 1950s. So, it should not be too surprising that the coherence and stability of fracterms is not matched by that of fracvalue. Fracterms are simply terms over a specific kind of signature (Definition \ref{def: fracterm}) and terms over signatures have been thoroughly analysed and applied in logic and computing. However, fracvalue depends on an infinite range of examples of number systems.  

In mathematics, numbers come in structures that likely have axiomatisations related to rings and fields \cite{vanderWaerden1970}. However, number systems also include ancient forms of numeration, such as the roman number system. The space of meanings for fracvalue presents ontological problems: understanding number becomes a philosophical task.

In order to organize how notions of number an be associated with fracvalue, we will think in terms of \textit{labels} and \textit{shapes}. 

Labels identify classes of numbers, notably: natural, integer, rational,  real, algebraic, transcendental, complex, quaternion, $p$-adic, nonstandard real, modulo$n$, floating point, fixed point, etc. Although the label carries important information and expectations, we need sharper resolution: a shape for a label provides a specific construction or way of defining a family of numbers for a label along with operations on these numbers. Thus, for each label there is a variety of shapes.

\begin{definition}\label{label&shape}
A {\em label} $L$ identifies a type of number. A {\em shape} $S$ for a label $L$ is a method for presenting the numbers of label $L$.  An {\em instance} of a shape $S$ is called an {\em $S$-number} for the number system with label $L$. For each label $L$ there is one or more shapes.

Let $i$ and $j$ be instances contained in a shape $S$, i.e., $S$-numbers for label $L$.  Then 

(i) $i =_S  j$ means that these instances are identical, i.e., $=_S$ is the native equality of the set of $S$-numbers of the shape $S$; and
\label{With or without subscritpt}

(ii)  $i \equiv_L j$ means that these $S$-numbers represent the same number in the number system with label $L$.
\end{definition}

Thus, for example, when we consider the natural numbers, we speak of a label $L=\mathsf{nat}$, and shape `$S$ = decimal'; hence, an  $S$-number becomes, say, a `decimal natural' or decimal $\mathsf{nat}$ (cf. ~\cite{Bergstra2020SACS}).

Clearly, we expect that for any shape for any label $L$, and any instances $i, j$ of the shape,
$$ i =_S j  \to  i \equiv_L j.$$
In fact, we are also familiar with the converse obtaining:

\begin{definition}\label{normal}
A shape for a label $L$ will be called {\em normal} if for any instances $i, j$ of the shape,
$$ i =_S j  \iff  i \equiv_L j.$$
If this equivalence fails the we will call the shape {\em subnormal}.
\end{definition}

 
\subsection{Examples of shapes for a number label}\label{example_shapes}

\noindent \textit{Number Label: $\mathsf{nat}$ for natural numbers.} Following Dedekind's algebraic analysis of whole numbers, a well-known shape for natural numbers is based on a unary operator of successor $S$: in symbols,
$$ 0, S(0), S(S(0)), S(S(S(0))),\ldots  .$$ We can refer to this shape as the Dedekind shape for naturals, and to its elements as Dedekind naturals.
Other well-known shapes are: 

- finite nonempty digit sequences with non-zero leading digit, together with $0$, ( i.e., decimal naturals without redundant leading zeroes, $L = \mathsf{sdn}$); 

- finite nonempty digit sequences (decimal naturals, $\mathsf{dn}$); this shape is subnormal, as for instance $007\equiv_\mathsf{dn}  7$;

- finite bit sequences (binary naturals); 

- finite hexadecimal sequences (hexadecimal naturals); 

- marks on a `number line'; 

-  von Neumann naturals:
$$0= \varnothing,  1 = \{ \varnothing \},  2 = \{ \varnothing ,\{ \varnothing \} \}, 
3 = \{ \varnothing ,\{ \varnothing \}, \{ \varnothing ,\{\varnothing \} \} \}, \ldots.$$

- Zermelo naturals: 
$$0= \varnothing,  1 = \{ \varnothing \},  
2 = \{ \{ \varnothing \} \}, 
3 = \{ \{\{ \varnothing \} \} \}, \ldots.$$

We refer to the various numbers as follows: Dedekind numbers are elements of the Dedekind shape. Binary numbers are numbers in the bit sequence shape(s), decimal numbers are numbers in the decimal shape(s), hexadecimal numbers are elements of the hexadecimal shape(s). For elements on a number line we have no specific naming (where number line shapes present naturals as equidistant points on a line), and von Neumann shape presents von Neumann numbers in the mentioned set theoretic fashion.
Von Neumann naturals are usually preferred over Zermelo naturals because (i) the definition of ordering is trivial, as it coincides with set membership $\epsilon$, and (ii) generalization to transfinite ordinals is quite natural, as taking the limit of a sequence of ordinals corresponds to the union thereof.
\\

\noindent \textit{Number Label: $\mathsf{int}$ for integer numbers.} Shapes of label $\mathsf{int}$ are usually made from numbers of a shape for label $\mathsf{nat}$. For integers, there are several well-known shapes. As a first example, the signed decimal naturals shape, uses pairs of the form $(+,a)$ and $(-,a)$ where $a$ is a natural taken from some chosen shape $S$ for label $\mathsf{nat}$ so that $a \neq_S 0$ and then takes these pairs together with $0$ as the domain of the shape (thus avoiding two zeros $(+,0)$ and $(-,0)$, which sometimes occur in computer arithmetics).

Another shape is to take equivalence classes of {\em difference pairs} $(a,b)$ with $a$, $b$ naturals taken from some chosen shape for label $\mathsf{nat}$, where equivalence is as follows: 
$$(a,b) \equiv_\mathsf{int} (c,d) \iff a+ d =_S c + b.$$

\noindent \textit{Number Label: $\mathsf{rat}$ for rationals.} The pair class shape plays a central role, based on some chosen shape $S$ for integers, which serves as a parameter for the construction. The pair class shape $PCS_{\mathsf{rat}}(S)$ is made out of equivalence classes pairs $(a,b)$, with $a, b$ taken from a shape $S$ for integers. Equivalence of pairs is defined by: 
$$(a,b) \equiv_\mathsf{rat} (c,d) \iff (b=_S 0 \wedge d =_S 0) \vee (b \neq_S 0 \wedge d \neq_S 0 \wedge a \cdot d =_S b \cdot c).$$
From this equivalence relation equivalence classes $[(a,b)]$ are formed.  The instances of the shape $PCS_{\mathsf{rat}}(S)$ are sets $[(a,b)]$. Note that this shape is normal in the sense of Definition \ref{normal}.

Another shape for rationals consists of {\em simplified simple fracterms}, shape, denoted $SSFT_{\mathsf{rat}}$, see Definition \ref{simple_fracterms}.

These differences matter as follows: in the pair class shape $PCS_{\mathsf{rat}}(S)$ rational numbers and fracterms are disjoint classes of entities, while in the $SSFT_{\mathsf{rat}}$ shape all numbers are also fracterms, though not all fracterms are numbers.

When communicating about, say, rationals different people may have different shapes in mind. That difference is usually immaterial and will not lead to confusion or complications because both persons are likely to use the same expressions for denoting numbers in their favoured shape of rationals.\\
\newline
\noindent \textit{Number Label: $\mathsf{real}$ for real numbers.} For the reals one often distinguishes three shapes: Cauchy sequences, Dedekind cuts, and infinitely long decimal expansions. If $\mathbb{R}$ is used in a text that may be read as: let $\real$ denote a fixed but arbitrary shape for the label $\mathsf{real}$.\\

\noindent \textit{Number Label: $\mathsf{complex}$ for complex numbers.} Two shapes are popular, both consisting of pairs of reals (i.e., elements of some shape for reals). In the complex plane shape $(a,b)$ is the interpretation of $a + i \cdot b$, and in the polar coordinates shape $(a,b)$ is the number serving as the meaning of $a \cdot e^{i \cdot b}$.


\subsection{Shapes and ontologies of numbers}
Thinking about the nature of numbers one is easily drawn to thinking about numbers in ways that are independent of their concrete symbolic and textual representations and of other constraints and factors associated with counting, measuring, etc. In abstracting from such details, great progress was made by Dedekind with his 1887 axiomatisation of the natural numbers, using constant  $1$, operation $x+1$, some axioms including the induction principle, and the concept of isomorphism: structures satisfying the axioms were all isomorphic \cite{Dedekind1887}. In much mathematical discourse, the details of data (= shape) disappear in structures that are characterised axiomatically and subsequently thought of `up to isomorphism'. 

Different normal shapes for the same label, when thought of as shapes for the same signature are supposed to be isomorphic, though there may be variations in particular with how partial functions are dealt with.  Thinking in terms of equivalence relations between mathematical entities, including the notions of equivalence given by isomorphisms of structures, is the standard route to finding commonality and invariant properties and defining abstractions.\footnote{This applies to elements, structures and classes of structures.} 
Using equivalence relations and equivalence classes  provides a well-trodden path towards discoursing on the ontology of numbers of a label. For \textit{elementary} arithmetic, however, this brand of structuralism is too complicated and working with particular shapes for each relevant label is preferable and necessary.


\subsection{Number as a cross-cutting notion}\label{cross-cutting_intro}

In arithmetic, there is no uniformity or consensus concerning the selection of shapes for the labels natural, integer, rational and real. The existence of different shapes for a label $L$ is seen as an advantage; however, it creates an obstacle that blocks us from simply speaking of the `numbers of label $L$'.  

Consider the label $\mathsf{nat}$ and contemplate shapes $S_1, S_2, \ldots, S_k$ for that label. Now, for any natural number, there will instances for that number in each shape:  $S_1$-numbers,  $S_2$-numbers, and so on. How can we obtain, say, `the number $7$' by way of abstracting from the particularities of shapes (abstracting from the decimal shape)?  

Now, we can view the natural number $7$ as a cross-cutting notion over the sample of available shapes for label $\mathsf{nat}$. In each of the shapes  $S_1, S_2, \ldots, S_k$ there is a counterpart of $7$. The collection of these counterparts, which may or may not contain the digit $7$, is the most abstract representation of $7$ given the sample of shapes  one is taking into consideration. In particular, in speaking of a cross-cutting notion we wish to make clear that \textit{no mathematical definition is given or sought for the concept of number, or for any specific numbers}. We return to this view later (in section \ref{fractions_revisited}).

The absence of such a definition is in line with the views proposed by Paul Benacerraf~\cite{Benacerraf1965}, who is sceptical about understanding individual numbers as well-defined and unique entities. When discussing or proving mathematical facts about naturals one must first select a shape, say $S_{i}$ for $\mathsf{nat}$, and then view or prove such facts \textit{purely as facts about the elements of the chosen shape $S_{i}$}.

That there are many shapes $S_1, S_2, \ldots, S_k$ for a label $L$ suggests that reductions and equivalences of shapes arise and need to be examined technically. 

\subsection{Example of a non-normal shape for $\mathsf{rat}$}

As explained above, defining rationals as classes of pairs of integers is a shape $PCS_{\mathsf{rat}}$ for the label $\mathsf{rat}$. In $PCS_{\mathsf{rat}}$, numbers are understood as equivalence classes of pairs. Notice that here if numbers are equal then they are the same as elements of a shape. The shape is normal.

There is an alternate way to consider numbers for the number label ${\mathsf{rat}}$: numbers \textit{are} the individual pairs, rather than the classes of pairs. This is a new shape for the label rational numbers that we refer to as the \textit{ratio-number shape} for rationals and denote by $RNS_{\mathsf{rat}}$.  This shape conceives a rational number as a ratio (rather than as a class of ratios). 

Now, $(1,2)$ and $(2,4)$ are \textit{not} the same instances of the shape in $RNS_{\mathsf{rat}}$, i.e., $(1,2) \neq_{RNS_\mathsf{rat}} (2,4)$ as numbers.  But, they are equal as as ratios and so as rational numbers in the conventional sense, i.e., $(1,2) \equiv_{\mathsf{rat}}  (2,4)$. The shape is not normal.  We can speak of $RNS_{\mathsf{rat}}$ as a \textit{subnormal shape} if the notions of equality and sameness of numbers are distinguished (Definition \ref{normal}).


\subsubsection{Algebra of ratio-numbers}
Let $S$ be some  arbitrary but fixed shape for integers.
We assume a sign function $\sg(-)$ on the $S$-integers with $\sg(p) = 1$ for positive $p$, $\sg(p) = -1$ for negative $p$, and $\sg(0) = 0$.
Let $\mathsf{RN}$ be the set of \textit{ratio-numbers} constructed by a function $(-,-): S \times S \to \mathsf{RN}$. We give it a familiar algebraic structure: the interpretation $\llbracket t \rrbracket$ for constants and functions is inductively defined as follows:

\begin{itemize}
\item $\llbracket 0 \rrbracket = (0,1)$, $\llbracket 1 \rrbracket = (1,1)$,
\item $\llbracket \mathsf{Num}(a,b) \rrbracket = (a ,1)$, $\llbracket  \mathsf{Denom}(a,b) \rrbracket = (b,1)$,
\item $ \llbracket -(a,b)  \rrbracket  = (-a , b)$,
\item $ \llbracket (a,b) \cdot (c,d) \rrbracket  = (a \cdot c, b \cdot d)$,
\item $ \llbracket (a,b) + (c,d) \rrbracket  = ((a \cdot c + b \cdot d), b \cdot d)$,
\item $ \llbracket \frac{(a,b) }{ (c,d)} \rrbracket  = \llbracket (a,b) \frac{1 }{ (c,d)} \rrbracket $,
\item $\llbracket \frac{1}{(a,b)} \rrbracket= (b,a \cdot \sg^2(b))$,
\item Instance equality: $\llbracket (a,b) =_\mathsf{RN} (c,d) \rrbracket \iff a =_S c \wedge b =_S d$,
\item Label equality: $\llbracket (a,b) \equiv_{\mathsf{rat}} (c,d) \rrbracket \iff (b =_S 0 \wedge d =_S 0)\vee (b \neq_S 0 \wedge d \neq_S 0 \wedge a \cdot d =_S b \cdot c $).
\end{itemize}

Using ratio-numbers with instance equality we obtain a new structure, which may be called a \textit{subnormal meadow} of (integer) ratio-numbers. 

In this structure, the operations $\mathsf{Num}$ and $\mathsf{Denom}$ do {\em not} respect equality, i.e., these implications {\em fail} in $RNS_{rat}$:
 $$x \equiv_{\mathsf{rat}} y \to \mathsf{Num}(x) =_S  \mathsf{Num}(y) \  \textrm{and}  \ x \equiv_{\mathsf{rat}}   y \to \mathsf{Denom}(x)=_S \mathsf{Denom}(y).$$ 
 
 What holds instead is 
 $$x =_\mathsf{RN} y \to \mathsf{Num}(x) =_S \mathsf{Num}(y) \  \textrm{and} \  x =_\mathsf{RN} y \to \mathsf{Denom}(x) =_S \mathsf{Denom}(y)$$
and then also 
$$ \mathsf{Num}(x)=_S \mathsf{Num}(y) \  \wedge \  \mathsf{Denom}(x)=_S \mathsf{Denom}(y) \to x \equiv_{rat} y .$$ 

In this structure, simplified integers that are ratio-numbers of the form $(a,1)$ can be distinguished from other integers,  i.e., ratio-numbers of the form $(a,b)$ with both $b\neq_S 0$ and $a$ an integer multiple of $b$.

Consider reading a fracsign as a ratio-number:  taking its numerator, by means of $\mathsf{Num}(-)$, may yield a result which is very different from the numerator found when reading the fracsign as a fracterm. For instance: if the shape for $\mathsf{int}$ is normal
 $$\mathsf{Num}(\frac{2}{\frac{4}{5}}) \equiv_\mathsf{int} 10\quad\mathrm{so~that}\quad \mathsf{Num}(\frac{2}{\frac{4}{5}}) =_S 10$$ 
 which we consider to be a counterintuitive outcome, given that upon reading $\frac{2}{\frac{4}{5}}$ as a fracterm one finds
 $$\mathsf{Num}(\frac{2}{\frac{4}{5}}) = 2.$$ 
 Therefore, the rational number shape is rejected as a semantic basis for elementary arithmetic. 
 \smallskip
 
\noindent \textbf{Remark.} Some dependance of $\mathsf{Num}(t)$ on the level assigned to $t$ will be impossible to avoid. Below we will ``accept'' that $\mathsf{Num}(\frac{2}{\frac{4}{5}})$ depends on the level of $\frac{2}{\frac{4}{5}}$: with level fracterm the result is $2$ while with level fracvalue the result is $\bot$, thereby expressing that \textit{fracvalues do not uniquely split into a numerator and a denominator}. 

We notice that yet another approach is to view $\mathsf{Num}(t)$ as the numerator of the rational $t$ written as a \textit{simplified} simple fracterm. The latter option leads to yet another  result:
$$\mathsf{Num}(\frac{2}{\frac{4}{5}})=_S 5.$$
We will for that reason also reject the third option as a definition for numerator and denominator in rational number arithmetic.


\subsection{Focus on normal shapes}
We will focus on normal shapes where equality is normal and no further refinement of equal entities is needed. Henceforth, by default, a shape is assumed to be normal; if not, the presence of a non-normal equality should be made explicit.

Working with a triple of fixed normal shapes for naturals, integers and rationals has the advantage that one may speak with some confidence of say  ``the number $2/5$''. Designing this triple of normal shapes in such a manner that naturals are also integers, and integers are also rationals, strengthens the intuition and identity of numbers. For details of such definitions of the respective shapes we refer to~\cite{Bergstra2020SACS}.


\section{The ontology of numbers as labels and shapes}\label{ontology}

To the best of our knowledge, the perspective and terminology of labels and shapes is new. Some of the intuitions have been discussed in~\cite{Bergstra2020SACS} though without making explicit the notions of labels and shapes and suggesting a general framework. 


\subsection{Label, shape and value}
In a practical context, a number is an element of a numerical data type, which is ``given'' somehow, made explicit, and agreed upon.  The organization of numerical data types takes place in three stages: labels, shapes (with variations), and values.

Labels primarily describe \textit{roles}. Naturals allowing addition and multiplication, and integers allowing subtraction are used for counting. Rationals allow division and are particularly important as they define the outputs of measurements, and the inputs for machines and devices. The reals encode geometries by means of infinitely close rational approximations of lengths and quantities; they host the constants $e$ and $\pi$, allow exponentiation and logarithm, and solve many equations, polynomial, differential, integral, etc. Note that these roles do not depend on any uniqueness of number systems---not even categoricity (= uniqueness up to isomorphism) is critical.

Numbers are distinguished in various ways as follows:

(i) {\em Number label}: natural, integer, rational real, complex. There are derived labels as follows: peripheral natural, peripheral integer, peripheral rational, peripheral real, peripheral complex. Typically,  $\bot$ may be introduced and understood as, say, a peripheral rational.

(ii) {\em Number shape}: for each label there are shapes---we have listed examples in section \ref{example_shapes}; number shapes include conventions on the presence of peripheral elements. Number shapes are arithmetical data types (or classes of arithmetical data types if parameters are left open). 

(iv) {\em Number value}: the actual value a number has is  some element of a given number shape for a given number label.

Confusingly, and seemingly in contradiction with apparent common sense, it is rather unhelpful to assume the existence of say natural numbers as a definite class of entities for which an agreed definition/description can be found. Thus, if in a text it is written that `$n$ is a natural number' then it is not presupposed that ``we already know'' what natural numbers are in some very general sense. Instead our idea is that natural indicates the choice of a label, and only \textit{after} choosing a shape for the label natural, and making that choice known to the reader of the text, do we have a class of entities in mind.

\subsection{Striving for numerical abstractions is a trap}
A formidable complexity arises from pursuing abstractions.  Different shapes, say $S_1$ and $S_2$, for the same label, say $L$, may be related by natural back and forth mappings, say
$$i_{1,2}\colon S_1 \to S_2  \ \textrm{and}  \  i_{2,1}\colon S_2 \to S_1.$$
For an element $a_1$ of shape $S_1$ more often than not there is a unique element $a_2$ that may serve as the $i_{1,2}(a_1)$ (and vice versa).
Perhaps, 
$$i_{2,1} (i_{1,2}(a_1) )= a_1$$
and for $k = 1,2$ one may also find that $a = i_{k,k}(a)$ for $a \in S_k$. 

Intuitively, for a series of shapes $S_1, S_2, \ldots $  one might consider the collection 
$$ \cup_ k \{ a_k= i_{1,k}(a_1) \ | \  a_1 \in S_1 \}$$
of all elements over all shapes $S_1, S_2, \ldots $ as the `abstract number' denoted by $a_1 \in S_1$. 

Now, in spite of this proposed approach to abstraction seeming tractable,  several awkward questions arise, such as: 
\medskip

(i) Can $a_k$ have any meaning without taking its host $S_k$ into account? 

(ii) What class is considered to play the role of $S_1$ and from which the seed element  $a_1$ chosen? 

(iii) How to deal with the situation that different shapes contain different forms of peripherals? 
\medskip

Solving these complex matters cannot plausibly be taken for a prerequisite for acquiring a minimum of arithmetical competence. The idea that the number concept must be understood in its most general way is a trap from which one may hardly recover.  However, the suggestion that the number concept is difficult and is a matter for academic mathematics and that its clarification is best delayed so that one may work with sloppy notions is also problematic.  A mathematician grown up with ZF set theory may have a different perspective on numbers than someone grown up in a (i) categorical framework, or (ii) an intuitionistic background or (iii) with some other constructive view of mathematics.

\subsection{Abstracting away from a single shape to a collection of shapes}

We feel that a reasonable picture can be found if one assumes that a person has in mind a possibly slowly growing \textit{portfolio of shapes} for a given label.  To begin with shapes for labels natural and integer, where some shapes are available in full detail and other shapes may be outlined imprecisely and merely indicators of a `kind of shape'. 

If it comes to acquisition of an abstract notion of number, say of label $L$, the advice is first to work with a single shape $S_1$ for label $L$. Working with $S_1$ involves making a connection between elements of $S_1$ and preferred syntactic expressions, and adopting rules of engagement for how to talk about closed expressions and values (in $S_1$) thereof. We assume that these conventions may be specific for $S_1$ even if the syntactic framework at hand is not specific for $S_1$.

Then, upon acquiring awareness of a second shape $S_2$ for label $L$, there should be an explicit thought process on how the availability of  $S_2$ may create a perspective from which adopting the orthodoxy that ``a number of label $L$ is an element of $S_1$'' is too narrow. Consider this idea:

The \textit{principle of incremental abstraction} names a process which may proceed in pace with incremental exploration of shapes. The idea is that given awareness of shapes $S_1,\ldots,S_k$ for $L$, one might acquire a more abstract view on numbers of label $L$ upon acquiring awareness of an additional shape $S_{k+1}$ for $L$.

We claim that one may safely work with shape $S_1$ without having made up one's mind on how to deal with incremental abstraction.
Equally unproblematic is it to change one's view on a preferred shape. If $S_2$ turns out to be more useful in practice then one may decide to work with $S_2$ and one may forget about $S_1$, if only temporarily.


\section{A naive form of structuralism}\label{structuralism}
An implication that emerges from the discussion of labels and shapes is this: there is no such thing as a precise concept of number in mathematics; this is not a new position. Number, as what we call a cross-cutting term,  presents aspects of Kramers's view. How do the intuitions of label and shape relate to foundational thinking about numbers?

Consider the idea that mathematical theories are fundamentally about abstract structures rather than the elements that make them up. This is a common view among pure mathematicians, and is called \textit{structuralism} by philosophers of mathematics. Structuralism has various technical and philosophical interpretations; an invaluable source for its history, methodology and philosophical intricacies is the compendium \cite{ReckSchiemer2020}.

In order to clarify the perspective on numbers in our paper, as a point of departure we follow the illuminating view of Erich Reck~\cite{Reck2003}. Reck analyses the position of Richard Dedekind, often seen as an originator of structuralism, and arrives at the conclusion that Dedekind represents what Reck calls \textit{logical structuralism}.


\subsection{Structure and shapes}

Consider natural numbers, labelled by $\mathsf{nat}$---similar considerations may be made about other numbers.
Assume we have available a collection of different shapes $S_1, S_2,..,S_k$ for the label $\mathsf{nat}$. A shape $S$ contains elements we call $S$-natural numbers. Both a shape and its elements are mathematical objects, and they have structures as understood by Dedekind.  The act of abstracting from unnecessary detail, which is central to Dedekind's views, is performed formally by thinking in terms of signatures, languages based on signatures, axioms and and isomorphisms between structures satisfying the axioms. Note that a signature provides a \textit{naming scheme} for constants, functions and relations of a structure---providing information beyond that of a similarity type.

For us, a collection of different shapes $S_1, S_2,..,S_k$ for the label $\mathsf{nat}$ gives rise to a cross-cutting notion of natural numbers, which as such, however, has no precise mathematical definition. 
Shapes for natural numbers come with substantial additional information: naming of some constants and functions, some expected properties of these constants and functions, and a connection between names and the entities that serve as interpretations of names.

The cross-cutting notion of number abstracts from details of the listed shapes, though not from details which these shapes have in common, if there are any. 

The suggestion is, for the case of elementary arithmetic, including school arithmetic,  to choose the decimal shape $S_{\mathsf{dec}}$ for natural numbers and to adopt the convention that $S_{\mathsf{dec}}$-naturals are simply referred to as naturals. Here the distinction between names and numbers is dropped: naturals are identified with names in a particular naming scheme. Reck~\cite{Reck2003} (p. 405) indicates that Dedekind, in unpublished work, made a similar suggestion, though with naturals constructed with a successor function (then named $\phi$) starting from a base element $1$ (rather than with decimals).

The idea is that a notion of natural number---i.e., a $S_{\mathsf{dec}}$-natural---results that suffices for educational purposes.  When needed, it may be further refined and adapted into the various forms of structuralism as listed by Reck: \textit{methodological structuralism}, \textit{set-theoretic structuralism}, \textit{ante rem structuralism}, and \textit{modal structuralism}, and \textit{logical structuralism} (following Dedekind, a position which has become less convincing, however, in view of the incompleteness results). 

We leave open the question to what extent there is a definition, conceptualization or identification of individual natural numbers as idealized abstractions  or understood in terms of the totality of conceivable shapes for natural numbers. Reck pays ample attention to the latter abstraction, which he identifies as a restriction to, and exclusive focus on, so-called \textit{constitutive} properties of numbers.

We assume that upon having several shapes for naturals in mind an urge to abstract from seemingly superfluous detail is created, though without any certainty that such an abstraction can be achieved, either by oneself or by other more competent scholars.\footnote{In a comparable manner inspection of say a range of automobiles of different brands may create the desire to find a most functional design of a car, without any guarantee that such a design can be found.}


\subsection{Formal foundations}\label{formal_foundations}

In a naive approach, elementary arithmetic is seen as a part of mathematics which is prior to set theory. Axiomatic set theory depends on a notion of formal syntax which is hard to imagine without the availability of natural numbers--- e.g., for counting numbers of arguments to a function, brackets, etc. Thus, axiomatic set theory is needed as a basis only after significant progress has been made in elementary arithmetic. Further, Peano arithmetic is seen as an axiomatic system which is prior to ZF(C) set theory. 

Now, non-axiomatic elementary arithmetic based on $S_{\mathsf{dec}}$-naturals is easily seen as being prior to Peano arithmetic. Once one adopts Peano arithmetic as a formal foundation for elementary arithmetic, numerals made with a successor notation may replace $S_{\mathsf{dec}}$-naturals as another (now   preferred) shape. Such numbers may be termed \textit{Dedekind-Peano naturals}. Upon making such a transition, one may drop the decimal shape and merely view decimal forms, like $05$, as \textit{names} or \textit{notations} for Dedekind-Peano numbers, like $SSSSS(0)$, rather than as numbers in their own right (i.e., in this case of the shape $S_{\mathsf{dec}}$-naturals). 

The scope of thinking about numbers with labels and shapes ought to include ancient and exotic number systems that have been replaced over time, notably the roman number system which is easily recognised as a system for naming natural numbers. The point is:

\textit{One's stock of shapes for a number label, here the number label $\mathsf{nat}$, may both grow and shrink in time.}

Viewing natural numbers as elements of the shape $S_{\mathsf{dec}}$ is compatible with having in mind that other shapes, taken from a stock of known shapes (e.g., the Von Neumann  naturals or the Zermelo naturals), for label $\mathsf{nat}$ may serve one's purposes as well. 

The use of `label' as an indication of a type of number is preferred over terminology such as  \textit{type, kind, sort, class, category, etc.} in order to express this design principle:
\smallskip

\textit{In giving a label, no assumption is made that numbers associated with the label have been yet defined or determined, further one is free to ignore and abandon any obligation to do so}. 
\smallskip

The situation is different with shapes, for speaking of a shape comes with an expectation that the shape can be clearly defined.

We use the term \textit{naive structuralism} as a description for the viewpoint outlined above. 
We opt for the qualification naive in view of the following reasons:

(i) the conventional preference for decimal notation; 

(ii) the fact that viewing $0$ and $1$ \textit{merely} as names for the numbers is unappealing for all (except logicians), indeed, it is hardly comprehensible for use in school arithmetic; 

(iii)  the explicit labelling of naturals for this shape as decimal naturals, 

(iv) adopting a form of relativism that is relative to one's stock of shapes for label $\mathsf{nat}$, thereby being an evolving position.

This naive realism may be considered a special case of `relative structuralism' as discussed in~\cite{ReckSchiemer2020b}. 

The suggestion that different notations give rise to different numbers can be dated back at least to 1867 through Hankel~\cite{Hankel1867}. For an exposition of Hankel's views see~\cite{Lawrence2021JAP}. So, we may perhaps speak of a naive Hankel-Dedekind realism for naturals---Hankel in view of the use of syntax, and Dedekind in view of using an instance structure, rather than some transcendental notion of number obtained by abstracting from unnecessary details. 

Using decimal naturals complies with Hankel's idea that \textit{numbers do not embody an intuition but have a primarily formal existence}.


\subsection{Avron on structuralism and signatures}
Avron's recent critique of structuralism for naturals, as outlined in~\cite{Avron2025}, consists of the following combination of views: 

(i) Naturals are elements of  a structure\footnote{In computer science terminology, a data type.} involving an initial element $\mathsf{init}$ and a successor function $\mathsf{succ}$.

(ii) Having just (i) in mind is insufficient for understanding naturals because it is then left undecided whether or not $0$ is viewed as a natural number; and as a consequence additional information is needed involving arithmetical operations able to distinguish the two cases: whether $\mathsf{init}$ is  $0$ and  $\mathsf{init}$ is $1$.

(iii) There is a preference for a structure whose elements have a `natural'  interpretation as cardinals thereby for  including a number $0$ as a natural number as that allows to quantify the empty set. 

(iv) These preferences in (iii) point in the direction of Von Neumann naturals, and the universal acceptance of Von Neumann naturals among mathematicians using ZF set theory is a significant fact. 

(We note that less unanimity is found for integers, yet less for rationals and that no unanimity is sought for reals for which several different shapes have survived the test of time.)

Avron provides convincing arguments for the following claim which we formulate, taking a small terminological liberty:

\begin{claim}
The notion of a signature enters the definition of natural numbers, merely working with $\mathsf{init}$ simply named as $0$ and successor does not suffice, a two-place function seems to be needed at least, such as addition and/or multiplication.
\end{claim}

The claim that this argument is valid matters because one of the two-place functions is necessary to distinguish $\mathsf{init}=0$ from $\mathsf{init}=1$. Using laws $0+0 =0$ and  $0+1 \neq 0$ one finds that with $\mathsf{init}=0$, $\mathsf{init} + \mathsf{init} = \mathsf{init}$  while with $\mathsf{init}=1$, 
$\mathsf{init} + \mathsf{init} \neq \mathsf{init}$. Alternatively if only multiplication is available one may use
 $\mathsf{succ}(\mathsf{init}) \cdot \mathsf{init} = \mathsf{init}$, true if $\mathsf{init}=0$, and  $\mathsf{succ}(\mathsf{init}) \cdot \mathsf{init} \neq \mathsf{init}$, $\mathsf{init}=1$, for making a distinction between both cases. So, abstraction from all detail (as advocated by Dedekind) fails on the objection of Avron that it will remain undecided where $0$ or $1$ stands at the root of the succession of natural numbers.


\section{Fractions revisited}\label{fractions_revisited}

Proceeding with the Hankel-Dedekind paradigm as suggested above we may now wonder which signature comes with a structural understanding of the rational numbers. The Hankel part suggest that having notations for numbers is helpful, so that having division as a part of the syntax becomes plausible. Now a question arises that is comparable with the issue about inclusion of $0$ in the natural numbers: is division a total function, and if so: what is $1/0$. 

In a naive Hankel-Dedekind approach to rational numbers, starting out with decimal integers, division is included in the signature so that the question `what is $1/0$' can be legitimately posed. We notice that taking division for a partial function induces complications as well, when it comes to conventions for informal arithmetic (see~\cite{BergstraT2024JLLI,BergstraT2025JLLI}).
Unlike the case with naturals and the inclusion of $0$, which has emerged as the more popular option, deciding upon the value of $1/0$ (or its lack of a value) turns out to be application dependent. When elementary education is the envisaged application our suggestion is that adopting $1/0 = \bot$ provides a workable option.

\subsection{Viewing fraction as a cross-cutting notion}\label{fraction_cross-cutting}

In the light of the previous discussion, we can clarify further what we have in mind by a cross-cutting notion, introduced  earlier, in Definition \ref{def:cross-cutting}.

Viewing each natural number as a cross-cutting notion that depends on the available stock of shapes for natural numbers (while taking $S_{\mathsf{dec}}$-naturals for naturals as a matter of convention) implements Benacerraf's scepticism about the feasibility of defining natural numbers without ruling out that option as a matter of principle. The situation with ``fraction'' is different, however.

In contrast, viewing fraction as a cross-cutting notion---ranging over fracvalue, fracterm, fracsign and fracsign occurrence--- comes with a scepticism concerning the option to find a proper mathematical or logical definition of fraction as a matter of principle. Indeed, it is not at all clear how to find a common abstraction for fracvalue, fracterm, fracsign, and fracsign occurrence, while at the same time it is hardly possible to conceptualize one of these notions without somehow having the other notions in mind as well.

The distinction between the uses of `cross-cutting' is worth examining. As a cross-cutting notion, fraction has certain properties that suggest explicit qualifications or identifications of (instances of) the notions are possible---fracvalue, fracterm, fracsign, and fracsign occurrence. Further these instances have degrees of abstraction that are discernible.  We propose to speak of fraction as a \textit{vertical cross-cutting notion}, whereas for the various number labels one may speak of \textit{horizontal cross-cutting} notions.

\begin{definition}
A {\em vertical cross-cutting notion} covers a range of specific notions which are ordered by level of abstraction. A {\em horizontal cross-cutting notion} covers a range of specific notions which are at the same level of abstraction.
\end{definition}

The higher levels of abstraction in the given range of abstractions may be thought of as the  philosophical rationale for the use of the lower levels of abstraction. The use of these lower 
levels is instrumental for dealing with higher levels of abstraction, by way of a reference, of calculation, or of physical representation.

By thinking in terms of $SSFT_{\mathsf{rat}}$-rationals (= simplified simple fracterms, based on decimal integers) room is left open for further abstraction which does away with the inessential details of $SSFT_{\mathsf{rat}}$-rationals. In other words the cross-cutting notion at hand is open ended, and leaves room for further evolution (or adaptation) if a person's views on the ontology of mathematical entities further develops.

Now, two open questions arise in this terminology of cross-cutting for which we have no answer yet, though we will make a guess in both cases:
\begin{question}
Can the use of horizontal cross-cutting notions be avoided?
\end{question}
We expect the answer to this question to be positive: horizontal cross-cutting notions may be made less ambiguous by choosing preferred entities, e.g.,shapes.
\begin{question}
Can the use of vertical cross-cutting notions be avoided?
\end{question}
We expect that the second question as a negative answer. We expect that vertical cross-cutting notions can be avoided in theory. However, in practice the use of vertical cross-cutting notions like fraction, product, sum and difference makes the natural language more efficient. Such efficiency is a fundamental advantage which cannot be missed in practice.


\section{Matters arising}\label{matters_arising}

A number of questions and directions arise from the analysis in this paper and here we draw attention to three: school teaching being a source of our interest in the ambiguity of fraction; the diversity of areas dependent on division; and formal logical theories upon which calculations with fracterms can be well-founded.


\subsection{Fractions and division by zero in school}\label{school}

Our discussion of fraction is centred on elementary arithmetic and on the rational numbers. Thinking of a fraction $\frac{x}{y}$ attracts a number of models of pedagogical interest. 

For example, in a \textit{separating or partitioning model} of division, in $\frac{x}{y}$ the denominator $y$ denotes the \textit{number of items or groups} used to separate or partition the whole: there are $y$ items or groups that make up the whole and $x$ is how many $y$'s  there are of some type. In a \textit{measurement model} of division, the denominator of $\frac{x}{y}$ denotes \textit{size of items or groups}: $y$ units make up the whole and $x$ is how many units  there are in the size of some subpart of the whole. Such models conjure up processes of organising or measuring. 

Now, if $y=0$  in these processes then
\smallskip

(a) partition: no groups are formed, no separation takes place---perhaps consistent with $\frac{x}{y} = 0$ or $\frac{x}{y} = x$;

(b) measure: the groups are empty of size 0---perhaps suggesting $\frac{x}{y}$ does not exist.
\smallskip

Either way the  process intuition stops.

Because fractions are so fundamental in elementary teaching, division by zero has some pedagogical interest, as do these subtle models of process; see the analysis in \cite{KajanderLovric2018,KajanderBoland2014}.

In~\cite{Bergstra2022TM}  we have discussed the fact that division may be characterised in terms of intentions, which has a role in managing the possibilities for a division by zero in evaluating an expression $P/Q$. In mathematics, one finds `retrospective division': mathematicians assume that the quotient of the values of $P$ and of $Q$ have been computed and now they merely look for a name for the quotient which is already known to exist.

Such a filter and guarantee is not feasible for division as it typically arises in executing a computer program.  `Prospective division' occurs in a computing machine where a command $x::= P/Q$ will express that as a next step the quotient of the values of $P$ and of $Q$ needs to be computed. The issue with prospective division is that it may be used in cases where it is not known---and there may be no way to know--- that $Q$ is non-zero. 

Such prospective division seems \textit{not} to play a significant role in teaching:  when considering elementary arithmetic in the way it often appears in school education, whenever division is required it becomes feasible to guarantee that division by zero won't happen. Thus, division by zero becomes a digression that seems not to be necessary and may not be made, unlike in the world of computing where it is unavoidable. So, the school case is a form of `retrospective division'.

Nevertheless, one may find a role for prospective division even in elementary arithmetic. Division may be included as a step in an informal algorithm, for instance consider the following algorithm $A_\mathsf{lin-eq}$:

``To solve the equation $a \cdot x = b$, use division and the result is $x = \frac{b}{a}$.'' 

Also a pupil may be tempted to apply $A_\mathsf{lin-eq}$ in case $a= 1$ and $b=0$.  This situation resembles executing a program  as the  machine is supposed to run a program independently of whatever are the effects of so doing.
A pupil exposed to division by zero ought to be encouraged to see what their pocket calculator does to $\frac{1}{0}$. Many will return $\frac{1}{0} = error$,  though some will return $\frac{1}{0} = 0$, albeit with a tag indicating an error. The pupil ought to be encouraged to compute further and see, say, $\frac{1}{0} + 1= error$, the point being that errors propagate.

Finally, let us observe that: 
\medskip

\textit{Elementary teaching of arithmetic is essentially the teaching of shapes, such as fracsigns and decimals, and rules for their transformation and identity. Said differently, elementary teaching of arithmetic is essentially the teaching of a calculus of (members of) shapes.} 
\medskip

Should one try to teach binary then a new topic is launched---its intimate connection with decimals and base $b>1$ number systems being obscure, if not invisible, to the pupil. 

Working with $\frac{x}{0} = 0$ is not uncommon, and has practical use in proof verification systems. We refer to this approach as Suppes-Ono fracterm calculus, in view of the original work of Suppes on division by zero (which yields zero) in~\cite{Suppes1957} (see also~\cite{AndersonB2021}), and the theoretical work done on the basis of this assumption in~\cite{Ono1983}.

Common meadows are algebraic fields that feature $\frac{1}{0} = \bot$, with $\bot$ a peripheral  element that propagate. We have developed algebraic and logical theories of these structures in~\cite{BergstraT2023}, and for the structure theory of the class of common meadows we refer to~\cite{DIAS2023,DIAS2024}. This approach to dealing with \textit{all} possible fracterms allows fracterm flattening (Definition \ref{flattening})
This feature we consider to be a necessity for any approach to elementary arithmetic. Without fracterm flattening, elementary arithmetic is in fact not anymore elementary!  In~\cite{BergstraT2022SACS} we have shown, that in practice the requirement of fracterm flattening brings with it the adoption of $\frac{1}{0} = \bot$, i.e., the circumstance that $\frac{1}{0}$ is propagates through formulae resulting in $\bot$.


\subsection{On ambiguity and probability}
Probability is an ambiguous word in mathematics. Probability either means {\em subjective probability} or it means {\em frequency based probability}. Although there may be many options in between, let us assume that probability is a word-bound hypothetical concept with these two major landmarks: subjective probability and frequentist probability, so probability is bi-ambiguous. 

Consider the ambiguity in terms of the notions in Section \ref{ambiguity_types}. Probability features significant ambiguity in a context where reporting of uncertain information is used for communication between proponents of frequentist probability and proponents of subjective probability. In many communities the very idea of subjective probability is either not known or not taken seriously, so that locality plays a role for the ambiguity of probability.   The ambiguity of probability is mostly non-controversial.

However, in forensic science, there seems to be (or have been) controversy on the matter with both positions having principled supporters. These issues are discussed in detail in \cite{Bergstra2019a}. As it turns out in the most basic case of Bayesian forensic reasoning both notions of probability occur in a single context, a situation which may give rise to confusion. On the one hand, the court will use subjective probability when arriving at a conclusion about the (probable) validity of a specific accusation.  On the other hand, the forensic laboratory may provide statistical information---based on work done in a frequentist mindset qua probability---which is turned into a likelihood (i.e., a conditional probability) or a {\em likelihood ratio}, which is then transferred to the court as novel information which might lead to an update of their subjective probability distribution about key events relevant to a case.

Likelihood ratio is a phrase made up from ratio which might inherit ambiguity from ratio as its component. We view likelihood ratio as non-ambiguous, however, assuming that a likelihood ratio is a number, which in the practice of approximate quantified reasoning is mostly a rational number.
In fact a likelihood ratio $r$ comes with the expression of which it is a value, which amounts to naming probabilities $\mathsf{A}$ resp. $\mathsf{B}$, without providing actual values, on the basis of which the likelihood ratio has been calculated.
It follows that there is room for an additional phrase: likelihood pair, a pair  $(p,q)$ with reals $p$ and $q$ such that  $p = \mathsf{P}(A \mid  B)$ and $q = \mathsf{P}(A \mid \neg B)$. 

A likelihood pair determines a unique likelihood ratio, while it provides additional information on top of its likelihood ratio consisting of a pair of conditional probabilities (which clearly cannot be retrieved from the mere value of its quotient). As non-numerical information a likelihood pair also includes the underlying fracterm which contains the names $A$ and $B$.


\subsection{Technical background: the algebra of division}

These reflections arise from our programme of technical research into new ways of specifying and reasoning about computer arithmetics. This uses the theory of abstract data types, which theorises how all forms of data can be accommodated in computer programs \cite{EhrichWL1997}. The theory has been developed from universal algebra \cite{MeinkeTucker1992}. 

A simple constraint that arises is the use of division $\frac{x}{y}$  in programs with the need to manage its partiality for $y=0$ in programming languages, computer algebra systems and theorem provers. Our approach to partiality has been to study semantic options for giving values to $\frac{x}{0}$, axiomatising their properties and analysing their effect on calculation and reasoning via formal calculi. For example, we have analysed in depth several techniques, including some popular hacks
$$\frac{x}{0} = 0$$
see, e.g., \cite{BergstraT2007,BergstraT2025_TCS}, and 
$$\frac{x}{0} = \bot,$$ 
where $\bot$ is a new element that behaves like an error value, which we call a peripheral, see \cite{BergstraT2023}. 

For computing, division with rational numbers is essential for defining numerical data, by measurement in terms of units and subunits, and for calculation, as all computer arithmetics are subsets of the rational numbers. From these theories, questions about the ambiguity of fractions, division by zero and the viability of using fractions for calculating arise, along with their implications for pedagogy.


\section{Concluding remarks}\label{Concluding_Remarks}

Ambiguities arise in all sorts of domains and can invite philosophical reflection. In science and mathematics, the value placed on analysis and precision make ambiguities particularly interesting to investigate.  In science, the classical domains---physics, chemistry, biology, etc.---offer different theoretical, experimental and practical worlds to explore concepts. In mathematics, one expects---might hope for---a simpler space with more technical commonalities and historical context. Here, we have applied the reduction method to the term `fraction' in the context of elementary arithmetic. In conclusion,  we reflect on the approach and outcomes.


\subsection{Summmary}
We introduced new terms to distinguish some of the meanings associated with fraction:  the concepts fracterm, fracvalue, fracsign and fracsign occurrence. Our basic idea is to use the term \textit{fracterm} to make precise the structural interpretation of fraction and \textit{fracvalue} for the numerical interpretation. With these aspects precisely defined, we can make the ambiguity of fraction explicit, by synthesising the notion of a \textit{fraxion} being the disjoint union of the four notions just mentioned.  In applying Method \ref{general_method}, these five notions are landmark notions which are more easily explained than the target notion of fraction; they specify different levels of abstraction in the presentation of certain mathematical entities. To conclude the application of the method, we discuss the effects of using these precise notions in arithmetical discourses; these discourses we call \textit{fractalk}.

As these ideas settle, we are then led to examine and clarify: \textit{what makes a number a fracvalue?} To do this we introduce notions of \textit{labels} and \textit{shapes} for number systems as new analytical concepts. A label identifies a kind of number, and a shape expresses the details of its textual presentation. Labels have many associated shapes. These fresh notions are explored and, later, are related to structuralist perspectives on numbers and their ontologies.


\subsection{Categories, concepts and notions}
In the discussion of labels and shapes we have not made use of category as a generalization of concept which admits ambiguity. The relations are thus: individual shapes, as well as elements therein, are concepts. Numbers and numbers systems are understood as crosscutting notions over a family of shapes for the same label are categories. The diversity of shapes for such (crosscutting) numbers and (crosscutting) arithmetical structures, viewed as categories, is a form of ambiguity. The latter ambiguity is non-controversial, with prominent shapes serving as landmarks. Given a finite sequence of shapes for a label, a disjoint union of these may be introduced as a way to find a more general disjunctive shape, comparable with the introduction of fraxions. However, working with such disjoint unions of shapes seems to bring no meaningful advantage.

Fraction is a category, and in the terminology of Brubaker~\cite{Brubaker2025}, fraction is a \textit{productively ambiguous category}. This means that different concepts serve as a meaning for fraction and each provide traction for the use of the word fraction, thereby reinforcing the usage of the category fraction, thanks to (rather than in spite of) its ambiguity. The phrase productive ambiguity is not new in the philosophy of mathematics, see, e.g.,Grosholz ~\cite{Grosholz2007} and Oliveri~\cite{Oliveri2011} where productive ambiguity is a rather technical notion different from our use of the phrase. Fraction has no definition, instead it has a description which amounts to  a plausible though non-exhaustive survey of  concepts which may serve as its meanings, including four major landmark notions: fracterms, fracvalues, fracsigns and fracsign occurrences.

For instance  viewing integer and fraction both  categories illustrates a certain degree of ambiguity in both case. However, both cases differ: the existence and even prominence of the shape of difference pairs (see~\ref{example_shapes} above)  for integers could have given rise to dedicated shape-dependent  terminology such as the positive part of an integer and the negative part of an integer, but such terminology has not been developed, let alone come into use.


\subsection{Canberra plan method}
An aim was to have studied ambiguity from first principles, by addressing some  mathematical raw material and observing what comes to the surface. The ambiguity of fraction arises in our technical work, and the Method \ref{general_method} is intended provide technical clarity; in particular, the method is intended to be naive and agnostic as to any philosophical positions and methods. Thus, how our observations in pursuit of clear meanings relate to particular philosophical approaches, old and new, is another task. At this stage we do not wish to place it in a framework that can accommodate ambiguity, e.g., \cite{Haueis2024}, or compare it with partially related studies \cite{Eagle2008}. 

However, the Informal Proposal \ref{landmarks} that introduces the paper sets the scene for the pursuit of a concept of `fraction', and the idea of seeking near approximations---landmarks---reminds us of the method of conceptual analysis nicknamed the \textit{Canberra Plan}. 

The Canberra Plan is a method designed to explore how concepts in one theory can be explained, refined or reduced to concepts in another. The original theories were physical theories but its philosophical discussion has wider scope. It arose in the work of Lewis \cite{Lewis1970}; a later development is Jackson \cite {Jackson1998}; see also a critique \cite{Raatikainen2021}.  The methods of the Plan involve the collection of basic statements about the concept which are called \textit{platitudes} and the construction of a Ramsey sentence to interpret it in a formal logic. 

We leave as an open question:  Can the method of the Canberra Plan be used on fraction and lead to similar conclusion as ours---namely that it qualifies as a category rather than a concept?

A platitude for 'fraction' might be `a fraction has a numerator and denominator'. The logic might be first or second order logic over the axiomatic theory of common meadows \cite{BergstraT2023,BergstraT2024c}.

Given our use of Brubaker's~\cite{Brubaker2025} ideas, which arose in the sociology of gender we note that the Canberra plan has been applied to gender in \cite{Podosky2026}, though without conclusive effect.


\end{document}